\documentstyle[12pt]{article}
\makeatletter
\newdimen\normalarrayskip              
\newdimen\minarrayskip                 
\normalarrayskip\baselineskip
\minarrayskip\jot
\newif\ifold             \oldtrue            
\def\arraymode{\ifold\relax\else\displaystyle\fi} 
\def\eqnumphantom{\phantom{(\theequation)}}     
\def\@arrayskip{\ifold\baselineskip\z@\lineskip\z@
     \else
     \baselineskip\minarrayskip\lineskip2\minarrayskip\fi}
\def\@arrayclassz{\ifcase \@lastchclass \@acolampacol \or
\@ampacol \or \or \or \@addamp \or
   \@acolampacol \or \@firstampfalse \@acol \fi
\edef\@preamble{\@preamble
  \ifcase \@chnum
     \hfil$\relax\arraymode\@sharp$\hfil
     \or $\relax\arraymode\@sharp$\hfil
     \or \hfil$\relax\arraymode\@sharp$\fi}}
\def\@array[#1]#2{\setbox\@arstrutbox=\hbox{\vrule
     height\arraystretch \ht\strutbox
     depth\arraystretch \dp\strutbox
     width\z@}\@mkpream{#2}\edef\@preamble{\halign \noexpand\@halignto
\bgroup \tabskip\z@ \@arstrut \@preamble \tabskip\z@ \cr}%
\let\@startpbox\@@startpbox \let\@endpbox\@@endpbox
  \if #1t\vtop \else \if#1b\vbox \else \vcenter \fi\fi
  \bgroup \let\par\relax
  \let\@sharp##\let\protect\relax
  \@arrayskip\@preamble}
\def\eqnarray{\stepcounter{equation}%
              \let\@currentlabel=\theequation
              \global\@eqnswtrue
              \global\@eqcnt\z@
              \tabskip\@centering
              \let\\=\@eqncr
              $$%
 \halign to \displaywidth\bgroup
    \eqnumphantom\@eqnsel\hskip\@centering
    $\displaystyle \tabskip\z@ {##}$%
    &\global\@eqcnt\@ne \hskip 2\arraycolsep
         $\displaystyle\arraymode{##}$\hfil
    &\global\@eqcnt\tw@ \hskip 2\arraycolsep
         $\displaystyle\tabskip\z@{##}$\hfil
         \tabskip\@centering
    &{##}\tabskip\z@\cr}
\makeatother

\def\beq{\begin{equation}}
\def\eeq{\end{equation}}
\def\bea{\begin{eqnarray}}
\def\eea{\end{eqnarray}}

\begin{document}

\begin{titlepage}
\begin{center}
\begin{flushright}{ NORDITA-93/20 \ \ FIAN/TD-04/93}\end{flushright}
\begin{flushright}{February 1993}\end{flushright}
On $p-q$ Duality and Explicit Solutions\\
in $c \le 1$ 2d Gravity Models\\[.4in]
{\large  S. Kharchev,}\\
\bigskip {\it  P.N.Lebedev Physics
Institute \\ Leninsky prospect, 53, Moscow, 117 924, Russia,
\footnote{E-mail address:
tdparticle@glas.apc.org}
}

\bigskip
{\large  A.Marshakov,}
\footnote{E-mail address: marshakov@nbivax.nbi.dk\ \
tdparticle@glas.apc.org}\\
\bigskip {\it  NORDITA,\\Blegdamsvej 17,
DK-2100\\Copenhagen {\O}, Denmark\\
and\\P.N.Lebedev Physics
Institute \\ Leninsky prospect, 53, Moscow, 117 924, Russia,
\footnote{permanent address}
}
\end{center}
\bigskip \bigskip

\begin{abstract}
We study the integral representation for the exact solution to
nonperturbative
$c \leq 1$ string theory. A generic solution is determined by two
functions $W(x)$ and $Q(x)$ which behaive at the infinity like $x^p$ and
$x^q$ respectively. The integral formula for arbitrary $(p,q)$
models is
derived which explicitly demonstrates $p-q$ duality of the minimal
models coupled to gravity. We discuss also
the exact solutions to string equation and reduction condition and
present several explicit examples.
\end{abstract}

\end{titlepage}

\newpage
\setcounter{footnote}0

\section{Introduction}

Recent years brought us to a great progress in
understanding of non-perturbative string theory. The key idea, established at
least for the most simple set of  $c \leq  1$  conformal theories interacting
with two-dimensional gravity, is the appearance of the structure of integrable
hierarchy in the description of generating function for physical correlators in
these models \cite{Douglas,FKN1}.

Fortunately, the particular solutions to non-perturbative string theory can be
singled from the whole set of solutions to the Kadomtsev-Petviashvili (KP) or
rather Toda lattice hierarchy by an additional requirement usually known in
the form of the
string equation, which allows one to present these particular solutions in a
form, based on the integral representations. In the papers
\cite{KMMMZ91a,KMMMZ91b,KMMM92a,KMMM92b,Mar92,KM92} it was shown, that there
exists
even a certain {\it matrix}-integral representation,
describing the particular subset of solutions to (reduced) KP-hierarchy
satisfying at the same time the string equation. The proposed matrix theory can
be considered ideologically as unifying theory for  $c \leq  1$  coupled to 2d
gravity string models
\footnote{and honestly for $(p,1)$ theories}
, allowing one to interpolate among them \cite{KMMMZ91a,KMMMZ91b,KMMM92b},
thus being a sort of effective string field theory \cite{Mar92}.

Below, we are going to investigate solutions to various $(p,q)$-models (with
central charges  $c_{p,q}=1-6{(p-q)^2\over pq} )$  coupled to $2d$ gravity in
more details. Moreover, we would stress the advantages of their integral (or
better multiple integral) representation, proposed in \cite{KM92} for the
particular ``stringy" solutions to KP hierarchy.

In particular, we are going to argue, that for higher critical points
the integral representation still makes sense, though it does not give
at the moment the final "string-field-theory" answer.
Shortly, higher critical points can be described using the same
``action" principle for the Douglas equations \cite{Douglas}, based on study
of the quasiclassical limit
\cite{Kri,TakTak}, but the exact answer has much more complicated form
and depends in general on {\it two} functions $W(x) = x^p + ...$ and
$Q(x) = x^q + ...$.
In contrast to the simple $(p,1)$ situation (with $Q(x) = x$),
the integral representation for these solutions besides
the ``action" functional has very complicated structure of the integration
measure.
Nevertheless, this integral representation obeys the basic property of $p-q$
duality in the spirit of \cite{FKN3} and might turn to be useful for studying
the exact solutions in various examples.

The sense of the $p-q$-duality is no longer a simple symmetry of the
theory. Indeed, there exists a kind of transformation, connecting the
solution to the $p-q$-problem with the solution to the $q-p$-problem
\footnote{similiar in a sense to the $N-k$ duality for the $SU(N)_k$
Yang-Mills theory (Nahm transformation for the instantons $etc$)}.

In sect.2 we are going to repeat the main results of \cite{KMMM92b} on
$(p,1)$
solutions and speculate on naive ``analytic continuation" to higher critical
points. In sect.3 we will formulate the general prescription and derive an
integral formula, valid in the case of arbitrary $(p,q)$-solutions.
In sect.4 we consider
$p-q$ symmetry in the formulation using the reduction condition and the
action of the Kac-Schwarz operator \cite{KSch,Sch}. Sect.5 contains
several examples of $c<1$ exact $(p,q)$-solutions and sect.6 -- some comments
on what is supposed to be a particular example of $c=1$ situation. In sect.7
we give several concluding remarks.

\section{Review of $(p,1)$ models}

First, we remind that the partition
function is defined \cite{KMMMZ91a,KMMMZ91b} as a matrix integral
\beq\label{1}
Z^{(N)}[V|M] \equiv  C^{(N)}[V|M] e^{TrV(M)-TrMV'(M)}\int
DX\ e^{-TrV(X)+TrV'(M)X}
\eeq
over  $N\times N$  ``Hermitean" matrices, with the normalizing factor given by
Gaussian integral
\beq\label{2}
C^{(N)}[V|M]^{-1} \equiv  \int \hbox{  DY }\ e^{-TrV_2[M,Y]},
$$
$$
V_2 \equiv \lim _{\epsilon \rightarrow 0}
{1\over \epsilon ^2}Tr[V(M+\epsilon Y) - V(M) - \epsilon YV'(M)]
\eeq
and  $Z$  actually depends on  $M$  only through the invariant variables
\beq\label{3}
T_k = {1\over k} Tr\ M^{-k}\hbox{, }  k\geq 1\hbox{  ;}
\eeq
moreover, if rewritten in terms of $T_k$,  $Z[V|T] = Z^{(N)}[V|M]$  is actually
independent of the size  $N$  of the matrices.

As a function of  $T_k$ \  $Z[V|T_k]$ is a  $\tau $-function of KP-hierarchy,
$Z[V|T_k] = \tau _V[T_k]$, while the potential  $V$  specifies (up to certain
invariance) the relevant point of the infinite-dimensional Grassmannian.

For various choices of the potential  $V(X)$  the model (\ref{1}) formally
reproduces various
$(p,q)$-series:  the potential $V(X) = {X^{p+1}\over p+1}$  can be associated
with the entire set of $(p,q)$-minimal string models with all possible
$q$'s.
In order to specify $q$ one needs to make a special choice of $T$-variables:
all  $T_k= 0$, except for  $T_1$ and  $T_{p+q}$ (the symmetry between $p$ and
$q$ is implicit in this formulation).

However, this is only a formal consideration. For the potential  $V(X) =
{X^{p+1}\over p+1}$  the partition function  $Z[V|T_k] = \tau _V[T_k] \equiv
\tau _p[T_k]$  satisfies the string equation which looks like

\beq\label{4}
\sum ^{p-1}_{k=1}k(p-k)T_kT_{p-k} + \sum ^\infty _{k=1}(p+k)(T_{p+k} -
{p\over p+1}\delta _{k,1}) {\partial \over \partial T_k} \log \ \tau _p[T] = 0
\eeq
$i.e$. $\tau $-function is defined with all Miwa times (\ref{3}) around zero
values
(in $1/M$ decomposition like in original Kontsevich model) with the only
exception - $T_{p+1}$ is shifted what corresponds obviously to $(p,1)$
model.
Thus, we see that the matrix integral gives an explicit solution only to
$(p,1)$ string models which must be nothing but particular topological matter
coupled to topological gravity.

Of course, we still have an opportunity for analytic continuation in string
equation, using the definition of Miwa's times (\ref{3}). We have to satisfy
the
following conditions:
\beq\label{5}
T_1 = x
$$
$$
T_2 = 0
$$
$$
...
$$
$$
T_{p+1} - {p\over p+1} = 0
$$
$$
T_{p+q} = t_{p+q} = fixed
$$
$$
T_{p+q+1} = 0
$$
$$
...
\eeq
which is a system of equations on the Miwa parameters  $\{\mu _i\}$,  $i =
1,...,N$. So, to do this analytical continuation one has to decompose the whole
set
\beq\label{6}
\{\mu _i\} = \{\xi _a\} \oplus  \{\mu '_s\}
$$
$$
T_k = {1\over k} TrM^{-k} = {1\over k} \sum ^N_{j=1}\mu ^{-k}_j = {1\over k}
\sum    \xi ^{-k}_a + {1\over k} \sum ^{N'}_{j=1}{\mu '}_j^{-k} \equiv
T^{(cl)}_k + T'_k
\eeq
into ``classical" and ``quantum" parts respectively. In principle it is clear
that we have now to solve the equations
\beq\label{7}
T^{(cl)}_k = {1\over k} \sum    \xi ^{-k}_a = t_{p+q}\delta _{k,p+q} -
{p\over p+1}\delta _{k,p+1}
\eeq
and this can be done adjusting a certain block form of the matrix  $M$
\cite{KMMMZ91b,Mar92}. However, in such a way we can only vanish several first
times,
and the rest ones can be vanished only adjusting correct behaviour in the limit
$N \rightarrow  \infty $. The most elegant way
\footnote{due to A.Zabrodin}
is to use the formula
\beq\label{8}
\exp  (- \sum ^\infty _{k=1}\lambda ^kT^{(cl)}_k) =
\lim _{K \rightarrow \infty }
(1 - {1\over K}\sum ^\infty _{k=1}\lambda ^kT^{(cl)}_k)^K = \prod  _a (1 -
{\lambda \over \xi _a})
\eeq
and then the solution to (\ref{7}) will be given by  $K$  sets of roots of the
equation
\beq\label{9}
\sum ^\infty _{k=1}\lambda ^kT^{(cl)}_k - K = t_{p+q}\lambda ^{p+q} -
{p\over p+1}\lambda ^{p+1} - K = 0
\eeq
Obviously, the eigenvalues  $\xi _a$ will now depend on the size of the matrix
$N = (p+q)K + N'$  through explicit  $K$-dependence  $(\xi _a \sim
K^{1/(p+q)})$  and we lose one of the main features of $(p,1)$
theories -- trivial dependence of the size of the matrix. Now
we can consider only matrices of {\it infinite} size and deal only with the
{\it infinite} determinant formulas.

That is why we call such way to get higher critical points as a formal one.
Below we will try to understand an alternative way of thinking, connected with
so-called $p$-times. Indeed, it was noticed in \cite{KMMM92b} that there
exists {\it a priori} another integrable structure in the model (\ref{1}),
connected with the time
variables, related to the non-trivial coefficients of the potential  V. As a
results, the cases of monomial potential  $V_p(X) = {X^{p+1}\over p+1}$  and
arbitrary polynomial of the same degree  $(p+1)$  are closely connected with
each other. The direct calculation  \cite{KMMM92b} shows
\beq\label{10}
Z[V|T_k] = \tau _V[T_k] =
$$
$$
= \exp \left( - {1\over 2}\sum    A_{ij}(t)(\tilde T_i-t_i)(\tilde T_j-t_j)
\right)  \tau _p[\tilde T_k- t_k]\hbox{  ,}
\eeq
where
\beq\label{11}
V(x) = \sum _{i=0}^p {v_i\over i} x^i
$$
$$
\tilde T_k = {1\over k}Tr \tilde M^{-k}\hbox{ ,}
$$
$$
\tilde M^p = V'(M) \equiv  W(M)\hbox{  ,}
$$
$$
A_{ij} = Res_\mu  W^{i/p} dW^{j/p}_+\hbox{ ,}
\eeq
where $f(\mu )_+$ denotes the non-negative part of the Laurent series $f(\mu )
=
\sum \ f_i\mu ^i$ and
\beq\label{S.1}
\tau _p[T] \equiv  \tau _{V_p}[T]
\eeq
-- is the $\tau $-function of $p$-reduction. The parameters  $\{t_k\}$  are
certain linear combinations of the coefficients  $\{v_k\}$  of the potential
\cite{DVV,Kri}
\beq\label{12}
t_k = - {p\over k(p-k)}Res\ W^{1-k/p}(\mu )d\mu
\eeq
Formula (\ref{10}) means that ``shifted" by flows along $p$-times (\ref{12})
$\tau $-function is
easily expressed through the $\tau $-function of  $p$-reduction, depending only
on the difference of the time-variables $\tilde T_k$ and $t_k$. The change of
the spectral parameter in (\ref{5})  $M \rightarrow  \tilde M$  (and
corresponding
transformation of times  $T_k \rightarrow  \tilde T_k)$  is a natural step from
the point of view of equivalent hierarchies.

The $\tau $-functions in (\ref{10}) are defined by formulas
\beq\label{13}
\tau _V[T] = {\det \ \phi _i(\mu _j)\over \Delta (\mu )}
\eeq
and
\beq\label{14}
{\tau _p[\tilde T-t]\over \tau _p[t]} = {\det
\hat \phi _i(\tilde \mu _j)\over \Delta (\tilde \mu )}
\eeq
with the corresponding points of the Grassmannian determined by the basic
vectors
\beq\label{15}
\phi _i(\mu ) = [W'(\mu )]^{1/2} \exp \left( V(\mu ) - \mu W(\mu )\right)
\int   x^{i-1}e^{-V(x)+xW(\mu )} dx
\eeq
and
\beq\label{16}
\hat \phi _i(\tilde \mu ) = [p\tilde \mu ^{p-1}]^{1/2} \exp \left(
-\sum ^{p+1}_{j=1}t_j\tilde \mu _j\right)  \int
x^{i-1}e^{-V(x)+x\tilde \mu ^p}dx
\eeq
respectively. Then it is easy to show that $\hat \tau _p(T)$ satisfies the
$L_{-1}$- constraint with {\it shifted} KP-times in the following way
\beq\label{17}
\sum ^{p-1}_{k=1}k(p-k)(\tilde T_k-t_k)(\tilde T_{p-k}-t_{p-k}) +
\sum ^\infty _{k=1}(p+k)(\tilde T_{p+k}-t_{p+k}){\partial \over \partial %
\tilde T_k} \log  \hat \tau _p[\tilde T-t] = 0
\eeq
where $t_i$ defined by (\ref{12}) are {\it identically} equal to zero for  $i
\geq
p+2.$

The formulas (\ref{10},\ref{17}) demonstrate at least two things. First, the
partition function
in the case of deformed monomial potential $(\equiv  polynomial$ of the same
degree) is expressed through the equivalent solution (in the sense
\cite{Shiota,Tak}) of the same $p$-reduced KP hierarchy, second -- not only
$t_{p+1}$ but all  $t_k$ with  $k \leq  p+1$  are not equal to zero in the
deformed situation. We will
call such theories as {\it topologically deformed $(p,1)$} models (in contrast
to {\it pure $(p,1)$} models given by monomial potentials  $V_p(X))$, the
deformation is ``topological" in the sense that it preserves all the features
of topological models we discussed above. Moreover, this ``topological"
deformation preserves almost all features of $2d$ Landau-Ginzburg theories
and from the point of view of continuum theory they should be identified
with the twisted Landau-Ginzburg topological matter interacting with gravity.

These topologically deformed $(p,1)$ models as we already said preserve all
properties of $(p,1)$ models. Indeed, according to \cite{FKN1} shifting
of first times  $t_1,...,t_{p+1}$ is certainly not enough to get higher
critical points. To do this one has to obtain  $t_{p+q} \neq  0$, but this
cannot be done using above formulas naively, because it is easily seen from
definition (\ref{12}) of $p$-times, that  $t_k \equiv  0$  for  $k \geq  p+2$.
To do
this we have to modify the above procedure and we are going to this in next
section.

\section{General description}

The above scheme has a natural quasiclassical
interpretation. Indeed, the solution to $(p,1)$ theories given by the partition
function (\ref{1}) can be considered as a ``path integral" representation of
the
solution to Douglas equations \cite{Douglas}
\beq\label{20}
[\hat P,\hat Q] = 1
\eeq
where  $\hat P$  and  $\hat Q$  are certain differential operators (of order
$p$  and  $q)$ respectively and obviously  $p-th$ order of  $\hat P$  dictates
$p$-reduction, while  $q$  stands for  $q-th$ critical point. Quasiclassically,
(\ref{20}) turns into Poisson brackets relation  \cite{Kri,TakTak}
\beq\label{21}
\{P,Q\} = 1
\eeq
where  $P(x)$  and  $Q(x)$  are now certain (polynomial) functions. It is
easily seen that the above case corresponds to the first order polynomial
$Q(x) \equiv  x$  and the $p$-th order polynomial  $P(x)$  should be identified
with  $W(x)\equiv V'(x)$  \cite{Kri}. Thus, the exponentials in
(\ref{1}), (\ref{15}) and (\ref{16}) acquire an obvious sense of action
functionals
\beq\label{22}
S_{p,1}(x,\mu ) = - V(x) + xW(\mu ) = - \int ^x_0dy\ W(y)Q'(y) + Q(x)W(\mu )
$$
$$
W(x) = V'(x) = x^p + \sum ^p_{k=1}v_kx^{k-1}
$$
$$
Q(x) = x
\eeq
and we claim that the generalization to arbitrary $(p,q)$ case must be
\beq\label{23}
S_{W,Q} =  - \int ^x_0dy\ W(y)Q'(y) + Q(x)W(\mu )
$$
$$
W(x) = V'(x) = x^p + \sum ^p_{k=1}v_kx^{k-1}
$$
$$
Q(x) = x^q + \sum ^q_{k=1}\bar v_kx^{k-1}
\eeq
Now the ``true" co-ordinate is $Q$, therefore the extreme condition of action
(\ref{23}) is still
\beq\label{24}
W(x) = W(\mu )
\eeq
having  $x = \mu $  as a solution, and for extreme value of the action one gets
\beq\label{25}
\left.S_{W,Q} \right |_{x=\mu } = \int ^\mu _0 dy\ W'(y)Q(y) =
$$
$$
= \sum ^{p+q}_{k=-\infty }t_k\tilde \mu ^k
\eeq
where  $\tilde \mu ^p = W(\mu )$  and
\beq\label{26}
t_k \equiv  t^{(W,Q)}_k = - {p\over k(p-k)}Res\ W^{1-k/p}dQ\hbox{  .}
\eeq
We should stress that the extreme value of the action (\ref{23}),
represented in the form (\ref{25}), determines the quasiclassical (or
dispersionless)
limit of the $p$-reduced KP hierarchy \cite{Kri,TakTak} with $p+q-1$
independent flows. We have seen that in the case of topologically deformed
$(p,1)$ models the quasiclassical hierarchy is exact in the strict sense:
topological solutions satisfy the full KP equations and the first basic vector
is just the Baker-Akhiezer function of our model (\ref{1}) restricted to the
small phase
space. Unfortunately, this is not the case for the general $(p,q)$ models: now
the quasiclassics is not exact and in order to find the basic vectors in the
explicit form one should solve the original problem and find the exact
solutions of the full KP hierarchy along first $p+q-1$ flows. Nevertheless, we
argue that the presence of the ``quasiclassical component" in the whole
integrable structure of the given models is of importance and it can give, in
principle, some useful information, for example, we can make a conjecture that
the coefficients of the basic vectors are determined by the derivatives of the
corresponding {\it quasiclassical $\tau $}-function.

Returning to eq.(\ref{26}) we immediately see, that now only for  $k \geq
p+q+1$ \ \ $p$-times are identically zero, while
\beq\label{27}
t_{p+q} \equiv  t^{(W,Q)}_{p+q} = {p\over p+q}
\eeq
and we should get a correct critical point adjusting all  $\{t_k\}$  with  $k <
p+q$  to be zero. The exact formula for the Grassmannian basis vectors in
general case acquires the form

\beq\label{dual}
\phi _i(\mu ) = [W'(\mu )]^{1/2} \exp ( - \left.S_{W,Q}\right |
_{x=\mu })  \int   d{\cal M}_Q(x)f_i(x) \exp \ S_{W,Q}(x,\mu )
\eeq
where  $d{\cal M}_Q(x)$  is the integration measure. We are going to explain,
that the integration measure for generic theory determined by two arbitrary
polynomials $W$  and  $Q$  has the form
\beq\label{29}
d{\cal M}_Q(z) = [Q'(z)]^{1/2}dz
\eeq
by checking the string equation. For the choice (\ref{29}) to insure the
correct
asymptotics of basis vectors  $\phi _i(\mu )$  we have to take  $f_i(x)$  being
functions (not necessarily polynomials) with the asymptotics
\beq\label{30}
f_i(x) \sim  x^{i-1}(1 + O(1/x))
\eeq

\section{$p$-reduction and the Kac-Schwarz operator}

To satisfy the string
equation, one has to fulfill two requirements: the reduction condition
\beq\label{31}
W(\mu )\phi _i(\mu ) = \sum  _j C_{ij}\phi _j(\mu )
\eeq
and the Kac-Schwarz \cite{KSch,Sch} operator action
\beq\label{32}
A^{(W,Q)}\phi _i(\mu ) = \sum    A_{ij}\phi _j(\mu )
\eeq
with
\beq\label{33}
A^{(W,Q)} \equiv  N^{(W,Q)}(\mu ){1\over W'(\mu )}
{\partial \over \partial \mu } [N^{(W,Q)}(\mu )]^{-1} =
$$
$$
= {1\over W'(\mu )} {\partial \over \partial \mu } - {1\over 2}
{W''(\mu )\over W'(\mu )^2} + Q(\mu )
$$
$$
N^{(W,Q)}(\mu ) \equiv  [W'(\mu )]^{1/2} \exp ( - \left.S_{W,Q}
\right |_{x=\mu })
\eeq
These two requirements are enough to prove string equation (see
\cite{KMMMZ91b} for details). The structure of action immediately gives us
that
\beq\label{34}
A^{(W,Q)}\phi _i(\mu ) = N^{(W,Q)}(\mu )\int   d{\cal M}_Q(z) Q(z)f_i(z)
\exp \ S_{W,Q}(z,\mu )
\eeq
and the condition (\ref{32}) can be reformulated as a $Q$-reduction property of
basis
$\{f_i(z)\}$
\beq\label{35}
Q(z)f_i(z) = \sum    A_{ij}f_i(z)
\eeq

Let us check now the reduction condition. Multiplying  $\phi _i(\mu )$  by
$W(\mu )$  and integrating by parts we obtain
\beq\label{35a}
W(\mu )\phi _i(\mu ) =
$$
$$
= N^{(W,Q)}(\mu )\int   d{\cal M}_Q(z) f_i(z)
{1\over Q'(z)} {\partial \over \partial z} [\exp \ Q(z)W(\mu )] \exp [-
\int ^z_0dy\ W(y)Q'(y)] =
$$
$$
= - N^{(W,Q)}(\mu )\int   d{\cal M}_Q(z) \exp [S_{W,Q}(z,\mu )] \left(
{1\over Q'(z)} {\partial \over \partial z} - {1\over 2} {Q''(z)\over Q'(z)^2} -
W(z) \right) f_i(z) \equiv
$$
$$
\equiv  - N^{(W,Q)}(\mu )\int   d{\cal M}_Q(z) \exp [S_{W,Q}(z,\mu )]
A^{(Q,W)}f_i(z)
\eeq
Therefore, in the ``dual" basis  $\{f_i(z)\}$  the condition (31) turns to be

\beq\label{35b}
A^{(Q,W)}f_i(z) = - \sum    C_{ij}f_j(z)
\eeq
with  $A^{(Q,W)} (\neq  A^{(W,Q)})$  being the ``dual" Kac-Schwarz operator
\beq\label{36}
A^{(Q,W)} = {1\over Q'(z)} {\partial \over \partial z} - {1\over 2}
{Q''(z)\over Q'(z)^2} - W(z)
\eeq

The representation (\ref{dual}), (\ref{29}) is an exact integral formula for
basis vectors
solving the $(p,q)$ string model. It has manifest property of $p-q$ duality (in
general $W-Q$), turning the $(p,q)$-string equation into the equivalent
$(q,p)$-string equation.

Now let us transform (\ref{dual}), (\ref{29}) into a little bit more explicit
$p-q$ form. As
before for $(p,1)$ models we have to make substitutions, leading
to equivalent KP solutions:
\beq\label{37}
\tilde \mu ^p = W(\mu )\hbox{, }    \tilde z^q = Q(z)
\eeq
Then we can rewrite (\ref{dual}) as
\beq\label{38}
\hat \phi _i(\tilde \mu ) = [p\tilde \mu ^{p-1}]^{1/2} \exp \left( -
\sum ^{p+q}_{k=1}t_k\tilde \mu ^k\right)  \int   d\tilde z
[q\tilde z^{q-1}]^{1/2} \hat f_i(\tilde z) \exp \ S_{W,Q}(\tilde z,\tilde \mu )
\eeq
where action is given now by
\beq\label{39}
S_{W,Q}(\tilde z,\tilde \mu ) =  - \left [ \int ^{\tilde z}_0d\tilde y
q\tilde y^{q-1}W(y(\tilde y)) \right ]_+  + \tilde z^q \tilde \mu ^p
$$
$$
= \sum_{k=1}^{p+q} \bar t_k \tilde z^k + \tilde z^q \tilde \mu ^p
\eeq
In new coordinates the reduction conditions are
\beq\label{40}
\tilde \mu ^p\hat \phi _i(\tilde \mu ) = \sum  _j
\tilde C_{ij}\hat \phi _j(\tilde \mu )
$$
$$
\tilde z^q\hat f_i(\tilde z) = \sum  _j \tilde A_{ij}\hat f_j(\tilde z)
\eeq
and for the Kac-Schwarz operators one gets conventional formulas
\cite{KSch,Sch,KMMMZ91b}
\beq\label{41}
\tilde A^{(p,q)} = {1\over p\tilde \mu ^{p-1}}
{\partial \over \partial \tilde \mu } - {p-1\over 2p} {1\over
{\tilde \mu ^p}}
 + {1\over p} \sum ^{p+q}_{k=1}kt_k\tilde \mu ^{k-p}
$$
$$
\tilde A^{(q,p)} = {1\over q\tilde z^{q-1}} {\partial \over \partial \tilde z}
- {q-1\over 2q} {1\over {\tilde z^
q}} + {1\over q} \sum ^{p+q}_{k=1}k\bar t_k\tilde z^{k-q}
\eeq
where for $(q,p)$ models we have introduced the ``dual" times:
\beq\label{S.6}
\bar t_k \equiv  t^{(Q,W)}_k = {q\over k(q-k)}Res\ Q^{1-k/q}dW
\eeq
in particularly, $\bar t_{p+q} = - {q\over p} t_{p+q} = - {q\over p+q}$ . Now
string equations give correspondingly
\beq\label{42}
\tilde A^{(p,q)}\hat \phi _i(\tilde \mu ) = \sum
\tilde A_{ij}\hat \phi _j(\tilde \mu )
$$
$$
\tilde A^{(q,p)}\hat f_i(\tilde z) = -\sum
\tilde C_{ij}\hat f_j(\tilde z)
\eeq

By these formulas we get a manifestation of $p-q$ duality if solutions to $2d$
gravity.

\section{Examples}

Now, let us
consider several explicit examples. First, for monomials  $W(x) = x^p$
and  $Q(x) = x^q$, $\tilde \mu  \equiv  \mu $,  $\tilde z \equiv  z$,
$\hat \phi _i \equiv  \phi _i$ and  $\hat f_i \equiv  f_i$, thus, the formulas
of the previous section will be
\beq\label{43}
\phi _i(\mu ) =  [p\mu ^{p-1}]^{1/2} \exp \left( - {p\over p+q}
\mu ^{p+q}\right) \times
$$
$$
\times  \int   dz [qz^{q-1}]^{1/2} f_i(z) \exp  \left( - {q\over p+q} z^{p+q} +
z^q\mu ^p\right)
\eeq
and the Kac-Schwarz operators acquire the most simple form
\beq\label{44}
A^{(p,q)} = {1\over p\mu ^{p-1}} {\partial \over \partial \mu } - {p-1\over 2p}
{1\over \mu ^p} + \mu ^q
$$
$$
A^{(q,p)} = {1\over qz^{q-1}} {\partial \over \partial z} - {q-1\over 2q}
{1\over z^q} - z^p
\eeq
For any $(p,q)$ theory with  $q>p$  the formula (\ref{43}) maps it onto the
corresponding ``dual" theory with  $q<p$  and vice versa.

In such way one can easily consider the $(p,1)$ topological theories as dual to
the "higher critical points" of the $(1,p)$ theory with the potential  $V_2(x)
=
{1\over 2}x^2$, $W_2 = x$. For this theory the ``topological" solution is
trivial
(for example, the partition function is given by a Gaussian integral and equals
to unity) so the basis vectors are
\beq\label{45}
f^{(1,p)}_i(z) = z^{i-1}
\eeq
and the Kac-Schwarz operator
\beq\label{46}
A^{(1,p)} = {\partial \over \partial z} - z^p
\eeq
preserves reduction of the corresponding $(p,1)$ model in a trivial way
\beq\label{47}
A^{(1,p)}f^{(1,p)}_i(z) = [ {\partial \over \partial z} - z^p] z^{i-1} =
$$
$$
= - z^{i+p-1} + (i-1)z^{i-2} = - f^{(1,p)}_{i+p-1}(z) + (i-1)f^{(1,p)}_{i-1}(z)
\eeq
In this particular case we see how the duality formula turns the problem of
finding nontrivial basis of \cite{KSch,KMMMZ91a,KMMMZ91b,K2} to the trivial
basis in
the Grassmannian (\ref{45}), corresponding to sphere.

In general case, we have no more the situation when a non-trivial problem
reduces to a trivial one. Moreover, it can be shown that for a generic
$(p,q)$ model the string equation reduces to an equation of generic
hypergeometric series
giving rise to (linear combinations of) generalized hypergeometric
functions \cite{Bailey,Bateman,Rainville}.

Indeed, we can obtain some particular solutions of the conditions (\ref{32})
as follows.
Let us consider the $(p,q)$ model with $q = pn + \alpha $, $\alpha  = 1$, ...,
$p-1$; $n = 0,1, 2, ...$ Using condition of $p$-reduction we can choose the
whole basis in the form
\beq\label{48}
\phi _{i+pk} = \mu ^{pk}\varphi _i\hbox{ , } i = 1\hbox{, ... , } p
\eeq
and therefore eq.(\ref{32}) give the system of equations for first $p$
vectors $\{\varphi _i(\mu )\}$ \cite{Sch}:

\beq\label{schwarz}
A\varphi _i =
\sum _{j} A_{ij}(\mu ^p) \varphi _{j}\hbox{ , }
i = 1\hbox{, ... , } p
\eeq
where in the case under consideration
\beq\label{50}
A = {1\over p\mu ^{p-1}} {\partial \over \partial \mu } - {p-1\over 2p}
{1\over \mu ^p} + \sum ^{pn+\alpha } kt_k \mu ^k \equiv
$$
$$
\equiv  N(\mu ) {1\over p\mu ^{p-1}} {\partial \over \partial \mu }
[N(\mu )]^{-1}
\eeq
and
\beq\label{50a}
N(\mu ) = [p\mu ^{p-1}]^{1/2} \exp \left( - \sum ^{p+q} t_k
\mu ^k \right)
$$
$$
q = pn + \alpha
\eeq
After the substitution
\beq\label{51}
\varphi _i =  N(\mu ) u_i(\mu )
\eeq
the system (\ref{schwarz}) acquires the form of

\beq\label{schwarz1}
{\partial \over \partial \mu ^p }u_i =
\sum _{j} A_{ij}(\mu ^p) u_j\hbox{ , }
i = 1\hbox{, ... , } p
\eeq

Now, let us present several explicit formulas for the simplest case of
 $p=2, q=2m-1$ solutions
\beq\label{matrix}
A_{ij}(\lambda) =
\left (
\begin{array}{cc} A_{11}(\lambda ) & A_{12}(\lambda )\\
A_{21}(\lambda ) & A_{22}(\lambda )
\end{array} \right )
\eeq
with $A_{11}(\lambda )$, $A_{22}(\lambda )$ and $A_{12}(\lambda )$ being the
polynomials of degree $m-1$ while the polynomial $A_{21}(\lambda )$ has
the degree $m$.

For the case of topological gravity $m=1$ (\ref{matrix}) looks like
\beq\label{m=1}
\left (
\begin{array}{cc} \beta & \alpha \\
\lambda + \gamma & - \beta
\end{array} \right )
\eeq
which can be by means of triangular transformations of the basis brought
to the form

\beq\label{m1}
\left (
\begin{array}{cc} 0 & \alpha \\
\lambda + \tilde \gamma & 0
\end{array} \right )
\eeq
and for the case of pure gravity ($m=2$) instead of (\ref{m=1}) and (\ref{m1})
one gets

\beq\label{m=2}
\left (
\begin{array}{cc}  m \lambda + b &  l\lambda + a \\
 a_2 \lambda ^2 +  a_1 \lambda + a_0 & - m
\lambda - b
\end{array} \right )
\eeq
which again, using the triangular transformations can be brought to the
form

\beq\label{m2}
\left (
\begin{array}{cc} \tilde  b &  l \lambda + a \\
a_2 \lambda ^2 + \tilde  a_1 \lambda + \tilde  a_0  & -
\tilde  b
\end{array} \right )
\eeq

Now, the eigenvalues of matrix (\ref{matrix}) given by
\footnote{implied by $Tr A = A_{11} + A_{22} = 0$, because as one can
check $Tr A$ gives contributions only to the {\it even} times in (\ref{rel})
and they can be easily eliminated by redefinitions with the help of
$p$-reduction conditions}

\beq
{\cal A}_\pm = - {1\over 2\lambda} \pm 2\lambda \left (
A_{11}^2 + A_{12}A_{21} + {A_{11} \over 2\lambda ^2} + {1\over 16
\lambda ^4} \right )^{1/2}
\eeq
are actually related to the "Krichever" times

\beq\label{rel}
{\cal A}(\lambda )_+ = \sum kt_k \lambda ^{k-1}
\eeq
which follows from the asymptotics of the solutions to the Kac-Schwarz
equations. It allows one to fix in (\ref{m1})

\beq
\alpha \sim t_3
$$
$$
\tilde \gamma \sim t_1
\eeq
while in (\ref{m=2}) and (\ref{m2}) the equations are more complicated
and even fixing $t_1 = 0 = t_3, \ \ t_5 = 2/5$ one ends up with a
nontrivial matrix

\beq\label{nmatr}
\left (
\begin{array}{cc} \tilde b &  \lambda +  a \\
\lambda ^2 -   a \lambda +  a^2  & -
\tilde b
\end{array} \right )
\eeq

Now the system of equations (\ref{schwarz}) with the matrix $A_{ij}$
given by (\ref{matrix}) can be diagonalized giving rise to

\beq\label{eq}
A_{12}u''_1 - A'_{12}u'_1 -
(A^2_{12}A_{21}+A'_{11}A_{12}-A_{11}A'_{12})u_1 = 0
\eeq
which for the case $m=1$ has as a solution

\beq\label{airy}
u_1(\lambda ) = \lambda \ _0F_1 \left [{4\over 3};-\lambda ^3 \right ] =
\lambda ^{1/2} J_{1/3}(2\lambda ^{3/2}) =
Ai(\lambda )
\eeq
while for the case $m=2$
\footnote{provided $a=0$, $\tilde b = 0$}

\beq\label{mcdon}
u_1(\lambda ) = \lambda ^2 \ _0F_1 \left [{7\over 5}; {1\over 5}\lambda ^5
\right ] = \lambda J_{2/5}(i{2\over 5}\lambda ^{5/2})
\eeq

The sense of these parameters, their relation to monodromy properties of
the solutions and relation to \cite{Losev} deserves further
investigation.

\section{Remarks on $c \to 1$ limit}

Let us now make some comments on $c=1$ situation. From
basic point of view we need in generic situation to get the most general
(unreduced) KP or Toda-lattice tau-function satisfying some (unreduced) string
equation. In a sense this is not a limiting case for $c<1$ situation but rather
a sort of ``direct sum" for all (p,q) models. This reflects that in conformal
theory coupled to $2d$ gravity there is, in a sense, less difference between
$c<1$ and $c=1$ situations than this coupling.

However, there are several particular cases when one can construct a sort of
direct $c \to 1$ limit and which should correspond to certain highly
``degenerate" $c=1$ theories. From the general point of view presented above
these are nothing but very specific cases of $(p,q)$ string equations, and
they could correspond only to a certain very reduced subsector of $c=1$ theory.

Indeed, it is easy to see, that for two special cases $p = \pm q$ the equations
(\ref{schwarz}) can be simplified drastically, actually giving rise to a single
equation instead of a system of them. Of course, these two cases don't
correspond to minimal series where one needs $(p,q)$ being coprime numbers.
However, we still can fulfill both reduction and Kac-Schwarz condition and
these solutions to our equations using naively the formula for the central
charge, one might identify with $c=1$ for $p=q$ and $c=25$ for $p=-q$.

Now, the simplest theories should be again with $q=1$. For
such case
``$c=1$" turns to be equivalent to a discrete matrix model \cite{KMMM92a}
while ``$c=25$" is exactly what one would expect from generalization of
the Penner approach \cite{Mar92,DMP}. Indeed, taking
{\it non-polynomial} functions, like
\beq
W(x) = x^{-\beta }
$$
$$
Q(x) = x^{\beta }
\eeq
the action would acquire a logariphmic term
\beq
S_{-\beta ,\beta } = - \beta logx + {x^{\beta } \over \mu^{\beta }}
\eeq
while equations (\ref{schwarz}) give rise just to rational solutions. It is
very
easy to see that $\beta = 1$ immediately gives the Penner model in the
external field,
which rather corresponds to ``dual" to $c=1$ situation with matter central
charge being $c_{matter}=25$ with a highly non-unitary realization of
conformal matter
\footnote{This $c=1$ \ -- \ $c=25$ duality might be also connected with the
known fact that there exists a Legendre transform between the
Gross-Klebanov solution to c=1 matrix model and the Penner model}.

On the other hand, $p=q=1$ solution is nothing but a trivial theory,
which however becomes a nontrivial discrete matrix model for unfrozen
zero-time. Moreover, these particilar $p = \pm q$ solutions become nontrivial
only if one considers the Toda-lattice picture with negative times being
involved into dynamics of the effective theory. On the contrary, we know
that $c<1$ $(p,q)$-solutions in a sense trivially depends on negative times
with the last ones playing the role of symmetry of string equation
\cite{KMMM92a}. It means, that we don't yet understand
enough the role of zero and negative times in the Toda-lattice formulation.

{}From the point of view of the duality formula one can, however, try to
identify $c=1$ situation with the {\it fixed point} of the duality
transformation (\ref{dual}), $i.e.$ to put $W = Q$ and $\phi _i = f_i$ in
(\ref{dual}):

\beq\label{fp}
\phi _i(\mu ) = [W'(\mu )]^{1/2} \exp ( - {1\over 2} W^2 (\mu ))
\int   dz [W'(z)]^{1/2} \phi _i(z) \exp \left ( - {1\over 2} W^2 (z) +
W(z)W(\mu ) \right )
\eeq

In is obvious that (\ref{fp}) has trivial solutions (for $W = x^2$) related
to the discrete matrix model, however, it is more interesting if there
are more sensible solutions for $c=1$ situation.

\section{Conclusion}

Let us make some conclusive remarks. We tried to present in
the paper the exact mechanism of transitions among different $(p,q)$ solutions
of non-perturbative $2d$ gravity in the framework of general scheme proposed in
papers \cite{KMMMZ91a,KMMMZ91b,KMMM92a,KMMM92b,Mar92}. We demonstrated that a
naive
analytic continuation in the space of Miwa parameters though correct formally
leads to certain practical difficulties in explicit description of higher
critical points even in trivial situation. Instead, we proposed
a concrete scheme, which allow one to shift ``classical" counterparts of the KP
times, determined by the coefficients of the potential and by the choice of
right variable.

The corresponding integral representation is a direct consequence of the action
principle and in principle can be interpreted as a certain field theory
integral with a
highly nontrivial measure. It obeys manifest $p-q$ symmetry which is evident
and restores equivalence in motion along naively two different $p$- and $q$-
directions. Moreover, the appearance of higher degrees of polynomials can be
obtained by transformation from the higher critical points of lower $p$ models.
Various examples demonstrate that in principle one might look for a
self-consistent multiple integral description of the generic $(p,q)$
models though in contrast to the $q=1$ situation this is still an open
question.

One might also find some other questions to be answered. Even in a dual to
topological $(p,1)$
series model there exists nontriviality after  $\alpha logX$  term (and
negative times terms) are added to the potential. For the  $p=1$  model this
gives rise to a separate interesting problem -- the discrete Hermitean matrix
model \cite{KMMM92a} and the question is about interpretation of such
generalizations of nontrivial theories.

The other question is  more deep understanding of generic $c = 1$ situation
(which is not reduced to particular ``degenerate" cases considered in sect.6)
and the role of negative times: symmetry between positive and negative times,
the ``dissappearing" of negative times in $c<1$ case etc.
It is also
quite interesting to study the quasiclassical limit of general $(p,q)$
solutions
and to compare them with topological theories. This might shed light to the
underlying topological structure of generic $(p,q)$ models.

All these problems deserves further investigation and we are going to return to
them elsewhere.

\bigskip
We are deeply indebted to J.Ambj{\o}rn, D.Boulatov, P.Di Vecchia, A.Losev,
A.Mironov,
A.Morozov, J.Sidenius
and
A.Zabrodin for illuminating discussions. The work of A.M. was supported
by
NORDITA.

\end{document}